\documentclass[a4paper,fleqn,usenatbib]{mnras}

\usepackage{newtxtext,newtxmath}

\usepackage[T1]{fontenc}
\usepackage{ae,aecompl}

\usepackage{graphicx}	
\usepackage{amsmath}	
\usepackage{amssymb}	
\usepackage{epsfig}

\title[QPPs with an unusual phase shift]{Quasi-periodic pulsations in a solar flare with an unusual phase shift}

\author[E. G. Kupriyanova et al.]{
	Elena G. Kupriyanova,$^{1, 3}$\thanks{E-mail: elenku@bk.ru (EGK)}
	Larisa K. Kashapova,$^{2}$
	Tom Van Doorsselaere,$^{3}$\newauthor
	Partha Chowdhury,$^{4, 6}$
	Abhishek K. Srivastava,$^{5}$,
	and Yong-Jae Moon$^6$
	\\
	$^{1}$Department of Radio Astronomical Research, Central Astronomical Observatory at Pulkovo of the RAS, \\ Pulkovskoe shosse 65/1, Saint Petersburg 196158, Russian Federation\\
	$^{2}$Institute of Solar-Terrestrial Physics SB RAS, Lermontova st. 126a, Irkutsk 664033, Russian Federation, email lkk@iszf.irk.ru\\
	$^{3}$Centre for mathematical Plasma Astrophysics, Mathematics Department, Celestijnenlaan 200B, bus 2400, Leuven B-3001, Belgium,\\ email tom.vandoorsselaere@kuleuven.be \\
	$^{4}$Engineering Science Department, University of Calcutta , 87/1 College Street, Kolkata-700073, India, email parthares@gmail.com\\
	$^{5}$Department of Physics, Indian Institute of Technology (BHU), Varanasi-221005, India, e-mail asrivastava.app@itbhu.ac.in \\
	$^{6}$School of Space Research, Kyung Hee University Yongin, Gyeonggi-Do, 446-701, South Korea, e-mail moonyj@khu.ac.kr
}

\date{Accepted XXX. Received YYY; in original form ZZZ}

\pubyear{2019}

\begin{document}
	\label{firstpage}
	\pagerange{\pageref{firstpage}--\pageref{lastpage}}
	\maketitle
	
	\begin{abstract}
		Two kinds of processes could occur during the flare decay phase: processes of energy release or processes of energy relaxation. Quasi-periodic pulsations (QPPs) of the broadband emission are a good tool for the verification of mechanisms.
		We aim to study the processes during the decay phase of the X-class solar flare SOL2014-03-29T17:48.
		The observations in X-ray, microwave, and extreme ultraviolet (EUV) bands are exploited to study the fine temporal, spatial, and spectral structures of the flare. The periods, amplitudes, and phases of both the fluxes and physical parameters (emission measure, temperature) are studied using standard methods of correlation, Fourier, and wavelet analyses. 
		It is found that the source of the QPPs is associated with the uniform post-flare loop. The X-ray source is located at the top of the arcade. QPPs with the similar characteristic time scales of $P\approx$~74--80 s are found in the X-ray (3--25~keV) and microwave (15.7~GHz) emissions. Besides, QPPs with the same period are found in the time profiles of both the temperature ($T_e$) and emission measure ($EM$).
		The QPPs in temperature and the QPPs in emission measure demonstrate anti-phase behavior.
		The analysis reveals the quasi-periodic process of energy relaxation, without any additional source of energy during the decay phase. The periods of the QPPs are in a good agreement with second harmonic of standing slow magneto-acoustic wave in the arcade which could be triggered by a Moreton wave initiated by the flare in the direct vicinity of the arcade.
		
	\end{abstract}
	
	
	\begin{keywords}
		Sun: flares -- Sun: corona -- Sun: X-rays, gamma-rays -- Sun: radio radiation -- Sun: UV radiation
	\end{keywords}
	
	
	\section{Introduction}
	
	Solar flares are one of the most energetic events in the solar atmosphere, which emit in the entire range of the electromagnetic spectrum from gamma rays to radio wavelengths. Emission flux increases by several orders of magnitude with an energy release up to $10^{32}$~ergs within few minutes of the impulsive phase of the flare. 
	Thereafter, the flux gradually decays to the quiescent level within a characteristic timescale. 
	
	The decay time depends on the mechanism in place. Two kinds of processes could occur during the flare decay phase: processes of energy release or processes of energy relaxation. The cooling time (in absence of additional sources of energy) is defined theoretically by both thermal conduction and radiative losses.  The characteristic decay timescales in this case vary from several tens of minutes to several hours depending on the physical conditions and on the loop length \citep{2006ApJ...638.1140J}. On the contrary, in some cases, evidence for the presence of the additional sources of energy were found in soft X-ray (SXR) and in hard X-ray (HXR) emissions. They are the long-lived cusp-shaped loop-top sources. The presence of a source of energy could prolong the decay phase up to from several hours to more than one day \citep[see, for example,][ and references therein]{2002ApJ...566..528I, 2011A&A...531A..57K}. Therefore, there is a need to identify which of the mechanisms operates in a given case. 
	
	A good tool for verification of the processes is hidden in the light curves of the flare emission. The light curves are frequently accompanied by periodic modulation of the emission, with sporadic fluctuations of the amplitude. This type of fine structures are called quasi-periodic pulsations, or QPPs \citep{2018SSRv..214...45M, 2016SoPh..291.3143V, 2009SSRv..149..119N}. QPPs were detected in all ranges of the electromagnetic spectrum, from radio to gamma rays \citep{1983ApJ...271..376K, 2010ApJ...708L..47N, 2010SoPh..267..329K, 2012ApJ...749L..16D, 2012ApJ...754...43S, 2015SoPh..290.3625S, 2016ApJ...833..284I, 2016A&A...585A.137G}. During strong solar flares, multiple layers of the solar atmosphere are involved in the process making it possible to do cross wavelength analysis of the emission. Parameters of the QPPs in different wavelength ranges are very sensitive to the variations of the parameters of the emitting plasma (temperature, density), the accelerated particles and the magnetic field. Therefore, QPPs are a good tool for diagnostics of both the physical parameters in the flare source and the mechanism. Particularly, 
	they allow to determine whether the observed brightness variations are the result of modulation of the acceleration process itself or by modulating the emission of already accelerated particles \citep{2016SoPh..291.3427K}.  
	
	The characteristic time\-scales, or periods, of the QPPs range from subsecond to several minutes, and sometimes they exhibit amplitude and period modulation \citep{2015A&A...574A..53K}. During the decay phase, the QPPs frequently appear as damped quasi-harmonic signals \citep{2016AdSpR..57.1456K, 2012ApJ...756L..36K}. The damping time is found to be proportional to the period of oscillations, and this proportion is similar for solar and stellar flares \citep{2016ApJ...830..110C}. The periods could evolve through their lifetime towards longer values \citep{2017ApJ...836...84D, 2016ApJ...827L..30H} or may remain stable \citep{2010SoPh..267..329K}. The damping oscillations are generally interpreted as relaxation of the eigen oscillations in the flaring loops excited by the flare, \textit{e.g.}, standing fast- or slow-mode waves \citep[for example,][]{2011A&A...534A..78R, 2012ApJ...756L..36K, 2013ApJ...778L..28S, 2015AdSpR..56.2769C, 2016ApJ...830..110C, 2016SoPh..291..877G}. The decay of the long-duration events is explained by the presence of additional sources of energy as plasma downflows \citep{2002ApJ...567L..85S} or magnetic reconnections \citep{2002ApJ...566..528I, 2011CEAB...35..115K}. The light curves of these events also exhibit quasi-periodic variations \citep{2011CEAB...35..103M}. However, the mechanisms behind the variations are not known yet.
	
	For example,  the flare SOL2014-03-29T17:48 was strong enough both to produce a white light flare \citep{2014ApJ...794L..23H} and to trigger a sunquake \citep{2014ApJ...796...85J} and Moreton wave \citep{2016SoPh..291.3217F}. Properties of this flare in different spectral ranges were intensively diagnosed by several authors \citep{2017SoPh..292...38W, 2016ApJ...827...38R, 2014ApJ...796...85J}.
	Analyzing this flare, \citet{2015ApJ...799..218Y} provide evidence of the upflow of hot plasma into the loop arcade. \citet{2015ApJ...804L..20A} have studied free magnetic energy evolution and found that its evolution is similar in coronal and chromospheric layers. However, no studies of the periodic properties of this flare were done. In this paper, we explore quasi-periodic processes during the decay phase of this solar flare. Analysing broadband emissions of the flare, we found an unusual phase behavior of the QPPs in the temperature ($T_e$) and emission measure ($EM$) time profiles. We discuss different scenarios of the QPPs to explain this.
	
	The paper is organized as follows. In Section~\ref{s:Observations} we describe observational data of the instruments {used in} this study and its analysis. Section~\ref{s:QPP} contains a brief outline of the method and results of the analysis of the parameters and sources of the QPPs. The principal mechanisms which could explain these observed periods and phase relationship are discussed in Section~\ref{s:Mechanisms}. Section~\ref{s:Conclusions} summarizes the principal results of the study.
	
	\section{Data of observations}\label{s:Observations}
	
\begin{figure*}   
		\centerline{
			\includegraphics[width=0.70\textwidth,clip=]{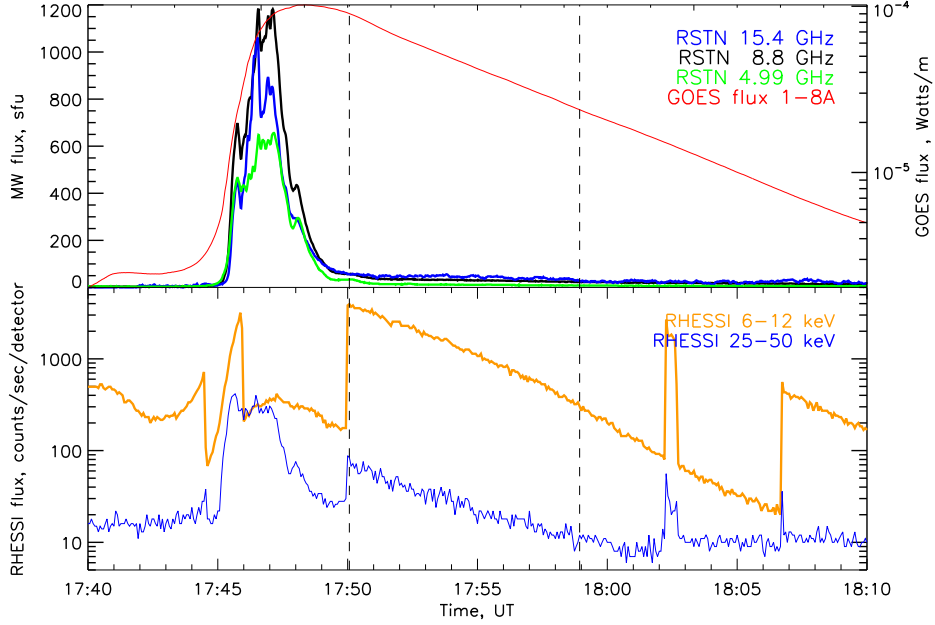}}
		
		\caption{The time profiles of X-ray and microwave emission during the flare. The top panel shows the comparison of the microwave emission by the RSTN and X-ray emission by GOES-15. The bottom panel shows the timeplots taken from RHESSI data. The vertical dashed lines marks the studied  time period. 
		}
		\label{f:timeplot1}
	\end{figure*}

	In this paper, we investigate quasi-periodic properties during the decay phase of the strong, X1.0 class flare. The flare SOL2014-03-29T17:48 occurred in active region NOAA 12017, located in the central part of the solar disc at the heliographic position N11W32 (Figure~\ref{f:timeplot1}). This flare started at 17:35~UT, reached its maximum at 17:48~UT, and then decayed and ended at 17:54~UT according to GOES observations. This event was simultaneously observed by a number of observatories. To detect QPPs during the decay phase of the flare, we have analyzed simultaneously the emissions in the X-ray range, in the microwave range, and in the Extreme Ultraviolet (EUV) range. The following data are used. 
	
	\subsection{X-rays}\label{s:ObservationsXRays}
	We used X-ray observations  of the flare by the \textit{Reuven Ramaty High Energy Solar Spectroscope Imager} (RHESSI) \citep{2002SoPh..210....3L}. As we can see in Figure~\ref{f:timeplot1} there were several changes of attenuators during the decay phase of the flare, resulting in large jumps of the data. The time interval 17:50--18:02 UT was without these jumps. Usage of the data of this period allowed us both to measure the localisation of the source and to analyse the time variability of the fluxes and plasma parameters. We processed the data   with the RHESSI software \citep{2002SoPh..210..165S} in order to construct the lightcurves, the images and  the spectral data. The spectra were obtained  with RHESSI's capability to carry out imaging spectroscopy. This approach helped us to avoid problems with subtraction of the background. The fit of the RHESSI spectrum allowed us to reconstruct the time profiles of the flare plasma parameters such as  the temperature ($T_e$), the emission measure ($EM$) and the spectral photon index ($\gamma$). For that purpose, we fitted the spectra with a model consisting of an optically thin thermal bremsstrahlung radiation function and a power-law function.
	\begin{figure}   
		\centerline{
			\includegraphics[width=0.45\textwidth,clip=]{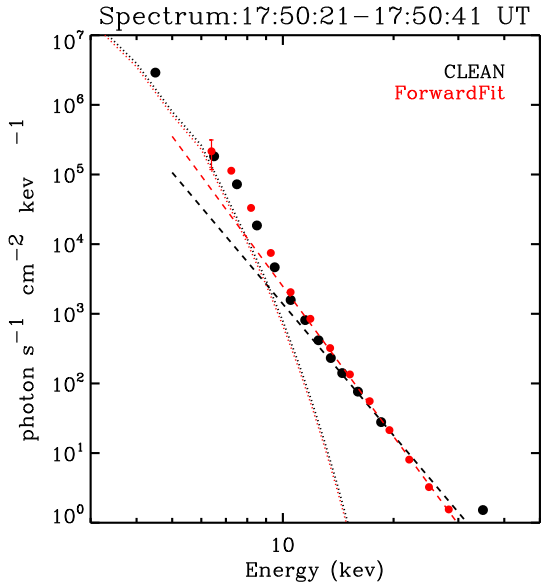}}
		
		\caption{ The spectra obtained with the imaging spectroscopy using the images reconstructed by CLEAN (the black colour) and ForwardFit (the red colour) algorithms. The results of fitting are shown by the corresponding colours. The thermal component of the fitting model is the dotted line and the power law component is the dashed line.	}
		\label{f:spect1}
	\end{figure}
	The minimal time cadence of RHESSI data is  4~seconds.  However, in this case the signal-to-noise ratio for the plasma parameters turned out too low. According to preliminary estimates, the QPP periods were more than 60 seconds, and therefore we used a binning over 20 second windows for the reconstruction of the images and the imaging spectroscopy. This cadence is appropriate for analysis of periods over 60~seconds and gives a sufficient signal-to-noise ratio for the plasma parameters in the flare source. We applied the \texttt{CLEAN} and \texttt{ForwardFit} (\texttt{FF}) algorithms for image reconstruction and compared the results of the spectral fitting for revealing possible instrumental effects. The comparison of the spectrum obtained using the images reconstructed by the different algorithms is shown on the Figure~\ref{f:spect1}. Both spectra were obtained by applying masks for getting the flux from the images. We use a circular mask with a radius of 100~arcseconds for the images reconstructed by the \texttt{CLEAN} algorithm and the mask radius was 33~arcseconds for the images obtained by the \texttt{FF} algorithm. One can see that the fluxes obtained by the different methods are in good agreement. We used  detector~9 (D9) for obtaining the spectra. This detector has the lowest resolution (180'') but the largest area.  However, data from the RHESSI detectors  D1, D3, D8 were also used both for image reconstruction and examination of the X-ray flux for various instrumental effects. The data from the detectors were processed, both integrated and separate for each detector. 
	
	The spectra allowed to separate energy bands and their time series were used for period analysis. We used the optically thin thermal bremsstrahlung radiation function as the thermal component and the single power-law function or the \texttt{thick2} algorithm for the description of the non-thermal component. The spectra obtained from the images were processed with standard methods using  the \texttt{SPEX} package \citep{2002SoPh..210..165S}, automatically taking into account the  albedo correction. This was necessary  because our event was located not far from the solar disk center. This simple model provided us electron temperature ($T_e$), emission measure ($EM$), electron flux variation and electron power-law index. Besides the mentioned time series, we construct the time series of the errors $\chi$ of the fitting to check if the QPPs are an artefact of the fitting or not. We found that the variation of electron flux and power-law index coincided with the  $\chi$  behaviour and concluded that the signal in these parameters  is a result of the fitting. According to spectra seen on the Figure~\ref{f:spect1}  the thermal component corresponded to energies up to $E < 12$--$15$~keV and the non-thermal component is definitely present for $E > 20$--$25$~keV.
	
	\subsection{EUV observations }\label{s:ObservationsEUV}
	In order to localize the source of the emission and to check its dynamic properties we performed comparative analysis of the HXR images with the images obtained by the \textit{Atmospheric Image Assembly} (AIA; Lemen et al. 2012) on-board the \textit{Solar Dynamics Observatory} (SDO). The AIA is a multichannel full disc imager with spatial resolution of 1.5~arcsec with a pixel size of 0.6~arcsec. Particularly, we checked the emission of the spectral lines 131{\AA} (Fe~\texttt{VIII}, Fe~\texttt{XXI}, Fe~\texttt{XXIII}) and 94{\AA} (Fe~\texttt{X}, Fe~\texttt{XVIII}). SDO/AIA provides the images with the best cadence time $\Delta t = 12.5$~s. This cadence is appropriate for analysis of periods longer than 40~s.
	
	\subsection{Microwaves}\label{s:ObservationsMW}
\begin{figure*}   
		\centerline{
			\includegraphics[width=0.70\textwidth,clip=]{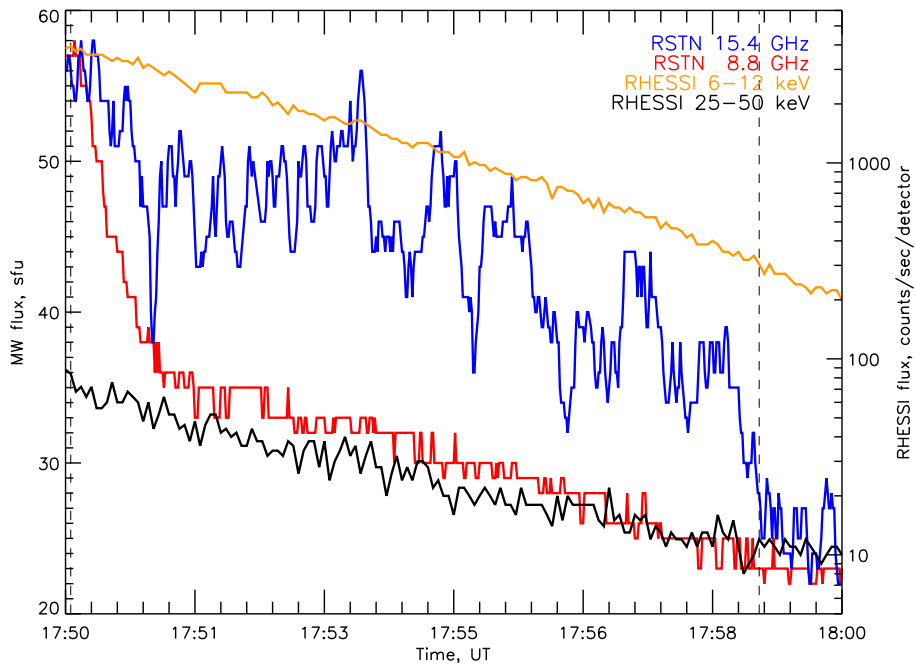}}
		\caption{The time profiles of X-ray and microwave emissions during the decay phase of the flare. The time interval under study is enclosed between two dashed vertical lines like those in Figure~\ref{f:timeplot1}. 
		}
		\label{f:timeplot2}
	\end{figure*}

	Microwave emission provides information about the accelerated electrons and about the emitting plasma. Comparison of the microwave of X-ray time profiles allows to judge about the source of the accelerated particles which produce both kinds of the emission. We use the full Sun observations of the ground-based \textit{Sagamore Hill Solar Observatory}, USA \citep{1981BAAS...13..400G}. This is one of the telescopes of the \textit{Radio Solar Telescope Network} (RSTN) which observes solar radio emissions at eight frequencies, 0.245, 0.41, 0.61, 1.415, 2.695, 4.995, 8.8 and 15.4~GHz. The telescope provides the temporal resolution $\Delta t = 3$~s which is more than enough to analyze the periods found in the flare under study. The flare was pronounced at frequencies higher than 4.99~GHz (Figure~\ref{f:timeplot1}). However, the pulsations are pronounced only in the time profile at 15.4~GHz (Figure~\ref{f:timeplot2}). So, we selected this frequency for the following analysis.  
	
	\section{Analysis of the parameters and sources of QPPs}\label{s:QPP}
	\subsection{Method of data processing}\label{s:QPPMethod}
	In this study we analyze the periods and phases of the QPPs. The first step before proceeding to the period analysis is the reduction of the original time series by subtracting a low-frequency trend. We have applied a smoothing with a running average over different time windows to define the trend. We subtract the trend from the original time series. This technique was applied earlier, {\textit e.g.} by \citet{2007A&A...473L..13O}, and described and tested by \citet{2010SoPh..267..329K} for white noise.
	As the results for various time windows are identical, in this paper we show the results only for a time window  of $\tau = 100$~s. The residual after subtraction is the high-frequency component. Subtraction of its average value makes it appropriate for the further periodic analysis.
	
	Alternatively, we apply the Fourier filtration method to define the high-frequency component independently \citep[\textit{e.g.}][]{2009A&A...493..259I}. 
	The results are compared for both methods. 
	The exceptions are the time series of $EM$, $T_e$, and $\chi$ have been obtained with $\Delta t = 20$~s. These time series are too short for applying the smoothing procedure. So, we extract their high-frequency component \textit{via} Fourier filtration only.
	
	We check the periodic properties of the time series by applying the standard Morlet wavelet transform \citep{1998BAMS...79...61T}. We consider white noise and red noise to calculate the confidence level for the detected periods. We have also calculated the global wavelet spectra of the data sets and compare it with the Fourier periodogram. 
	
	To analyze the phase properties of time series we use the high-frequency component extracted using the method of Fourier filtration. This method is preferable because it extracts the high-frequency component keeping the original phase of the pulsations. On the contrary, smoothing methods could distort the phase of a signal. We use a Fourier filter from 55~s to 125~s zeroing Fourier powers outside the filter.
	
	The phase relationship $\Delta \varphi$ between the oscillations in two different time series are analyzed using two standard methods: cross-correlation analysis and Fourier transform. For the first method, a cross-correlation function between two time profiles is calculated. The phase delay is a time lag between the maximum of the cross-correlation function and the zero time lag. This method is widely used in solar coronal seismology \citep[see, for example,][]{2013A&A...560A.107A, 2002A&A...385..671F}. In the second method, the phase is calculated based on the fast Fourier transform \citep[for example,][]{2002A&A...385..671F}. The Fourier frequency spectrum $z$ at frequency $\nu$ is a complex number $z_\nu = Re(z_\nu) +i Im(z_\nu)$. The phase at the selected frequency $\nu$ is determined by $\varphi_\nu = \arctan \frac{Im(z_\nu)}{Re(z_\nu)}$. So, the phases are calculated for the two considered time series and then the phase difference between them is found.
	
	\subsection{Periods}\label{s:Periods}
\begin{figure*}   
		\centerline{
			\includegraphics[width=0.50\textwidth,clip=]{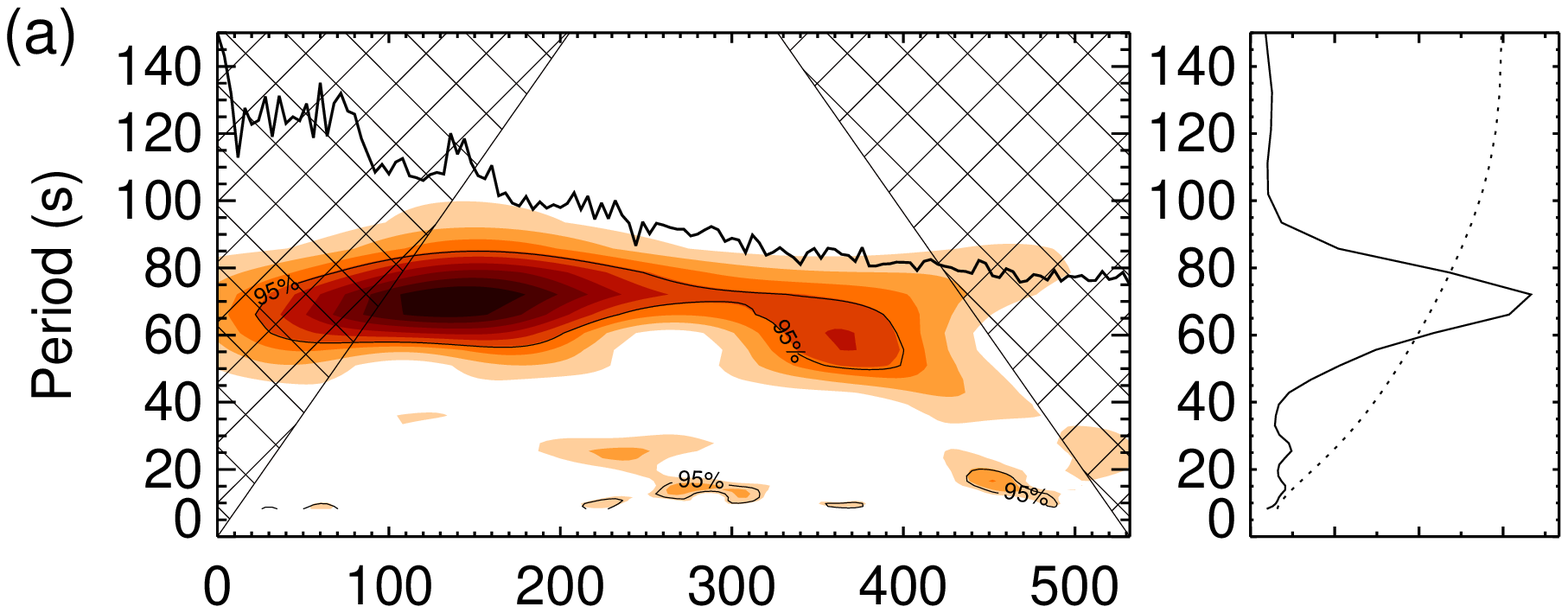}
			\includegraphics[width=0.50\textwidth,clip=]{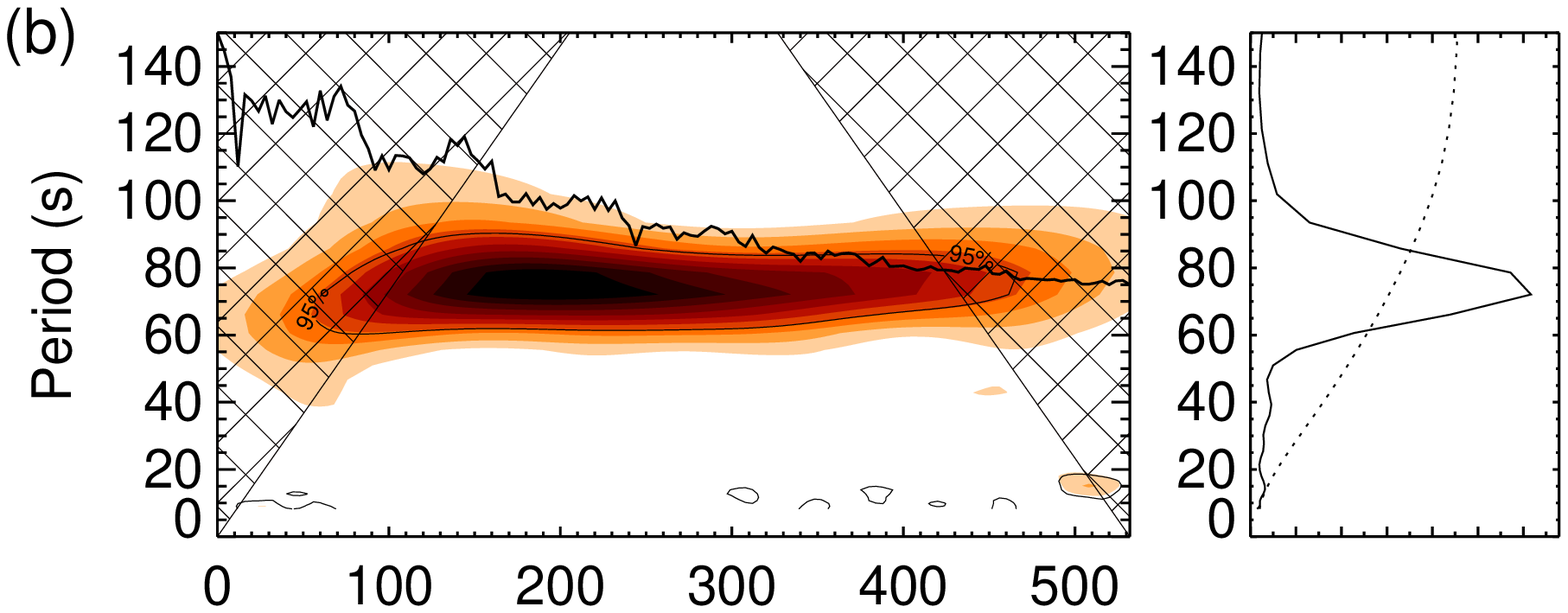}
		}
		\centerline{
			\includegraphics[width=0.50\textwidth,clip=]{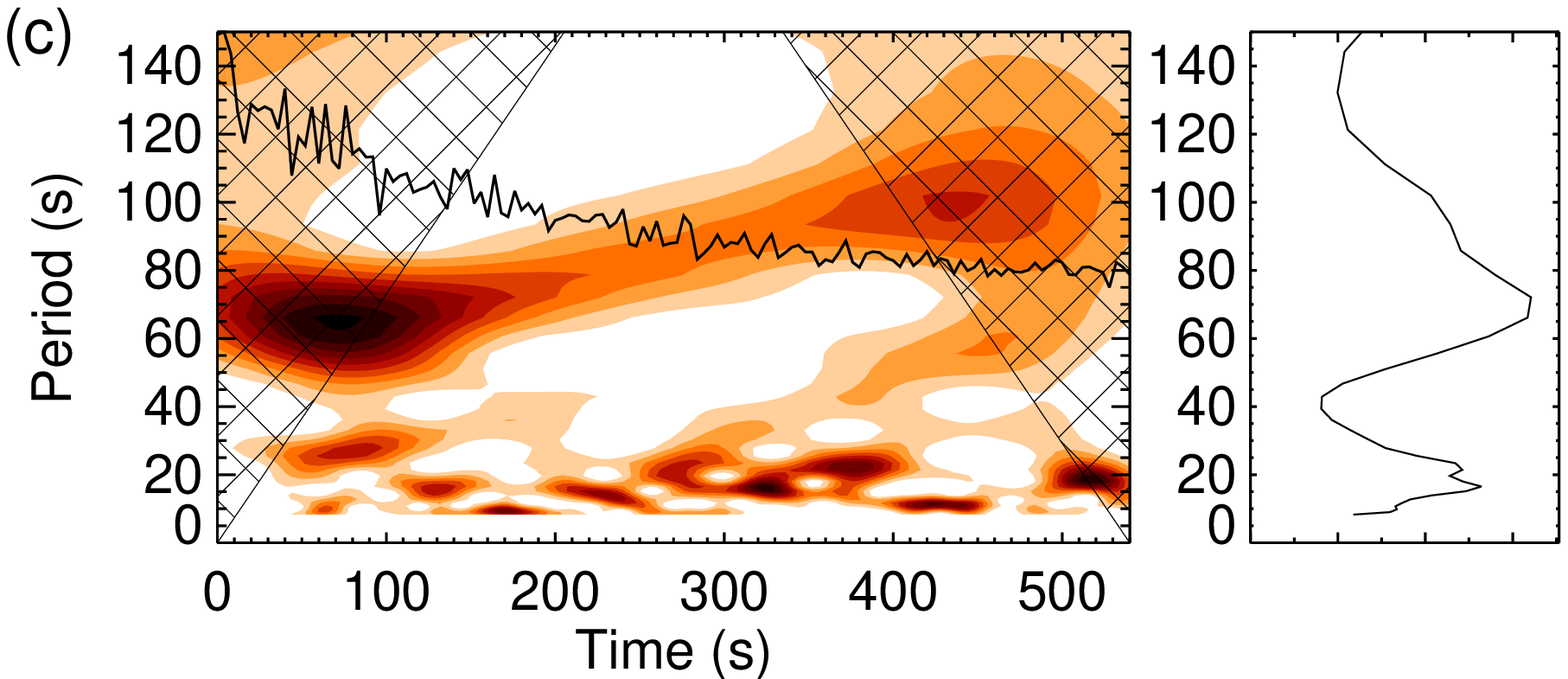}
			\includegraphics[width=0.50\textwidth,clip=]{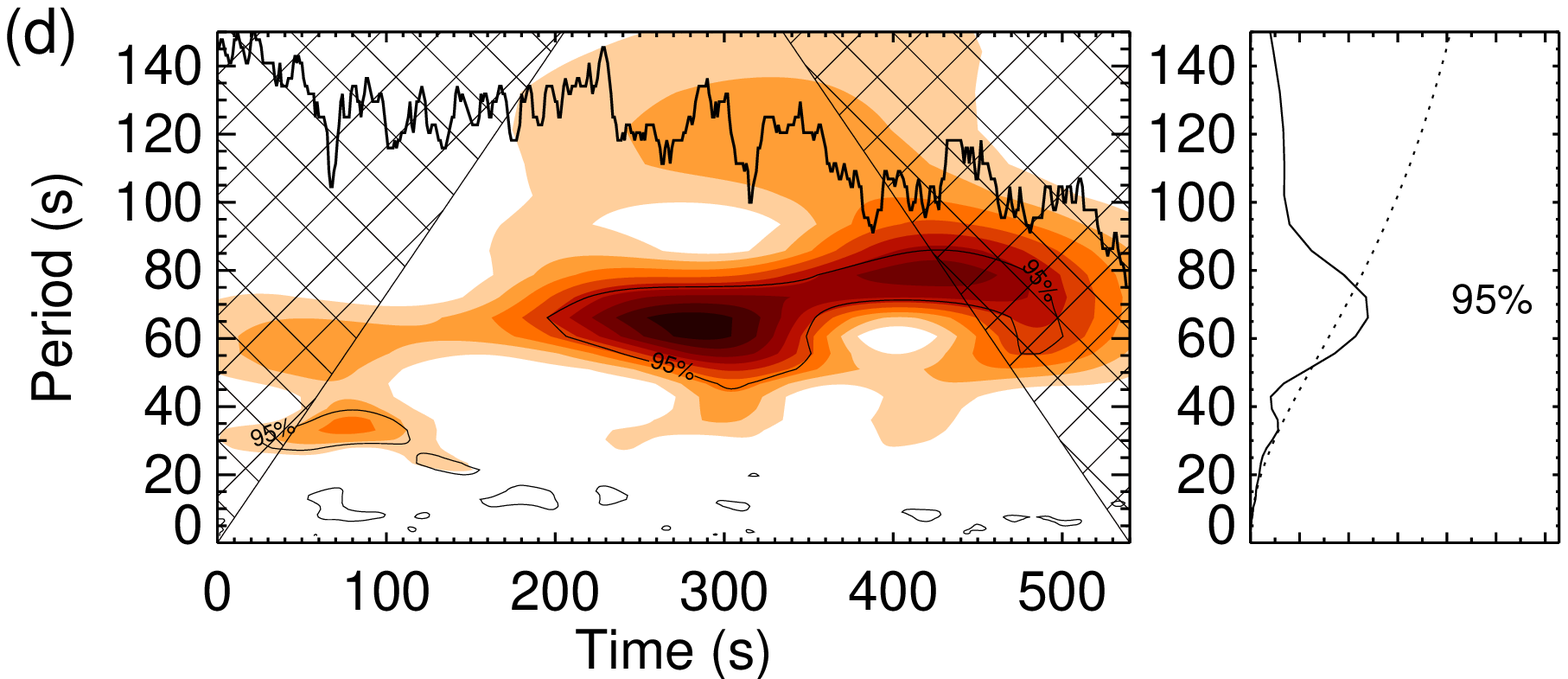}
		}
		\caption{
			Wavelet power spectra of the following time series:
			(a) the RHESSI flux at 3--6~keV, 
			(b) the RHESSI flux at 6--12~keV, 
			(c) the RHESSI flux at 25--50~keV, 
			(d) the RSTN flux at 15.4~GHz.
			The time axis corresponds to the interval 17:50:00--17:59:00~UT. The cadence time is $\Delta t = 4$~s. See Section~\ref{s:Periods} for a detailed description.}
		\label{f:wavelet1}
	\end{figure*}

	Figure~\ref{f:wavelet1} shows the results of the wavelet analysis of the time series of the hard X-ray in four energy bands (panels (a)--(c)) and microwave emissions (panel (d)). Note that we have analyzed the RHESSI data both integrated over all the detectors and obtained separately by detectors D1, D3, D8, and D9. Significant variations of the time profiles with a period $P \approx 72$~s are found both in the integrated data and in the time profiles obtained by the separate detectors D8 and D9, but no significant periods were found for the detectors D1 and D3. The wavelet spectra for D8 and D9 are similar within each energy band. So, we plot in panels (a)--(c) the wavelet spectra for the less noisy detector D9 only. 
	
\begin{figure*}   
		\centerline{
			\includegraphics[width=0.50\textwidth,clip=]{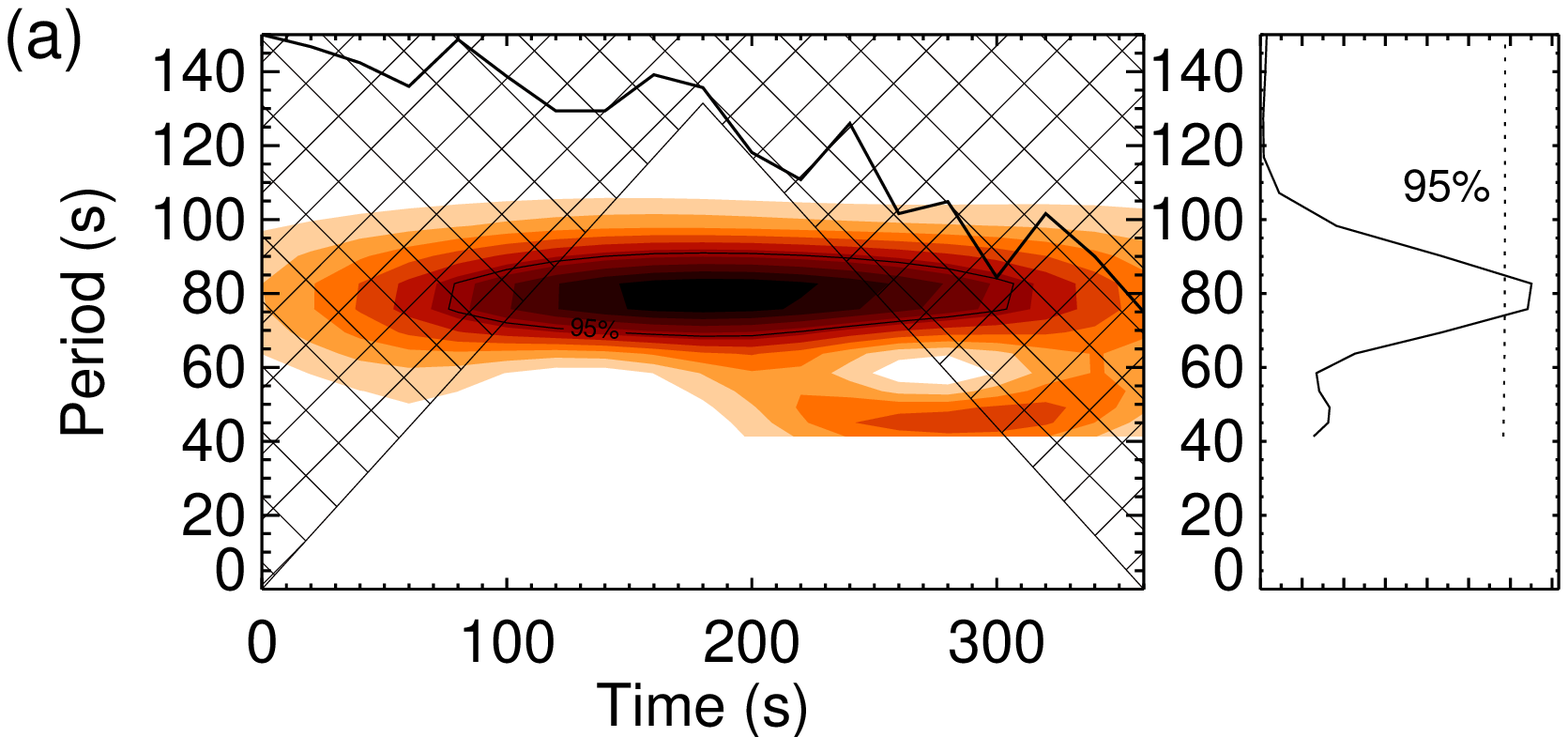}
			\includegraphics[width=0.50\textwidth,clip=]{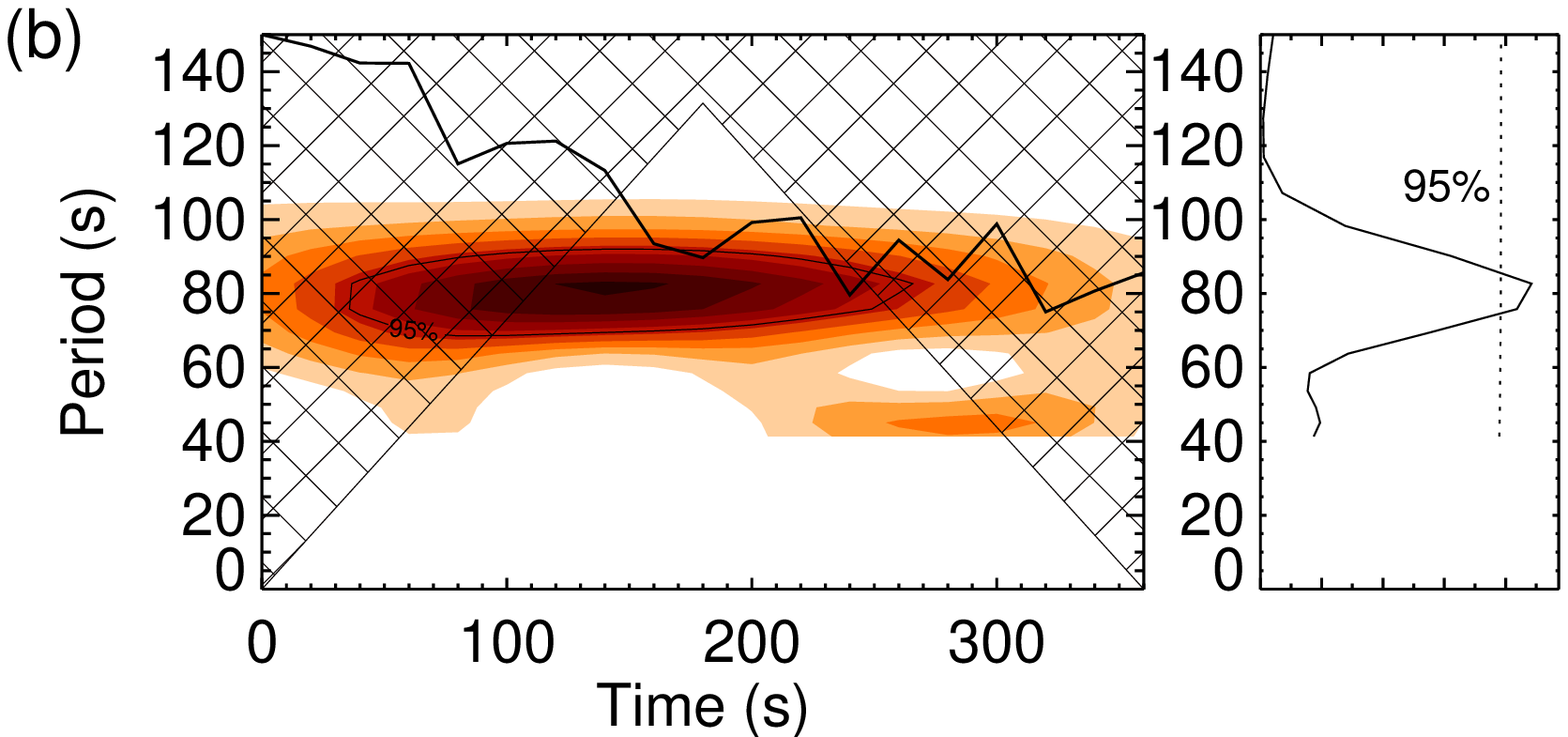}
		}
		\centerline{
			\includegraphics[width=0.50\textwidth,clip=]{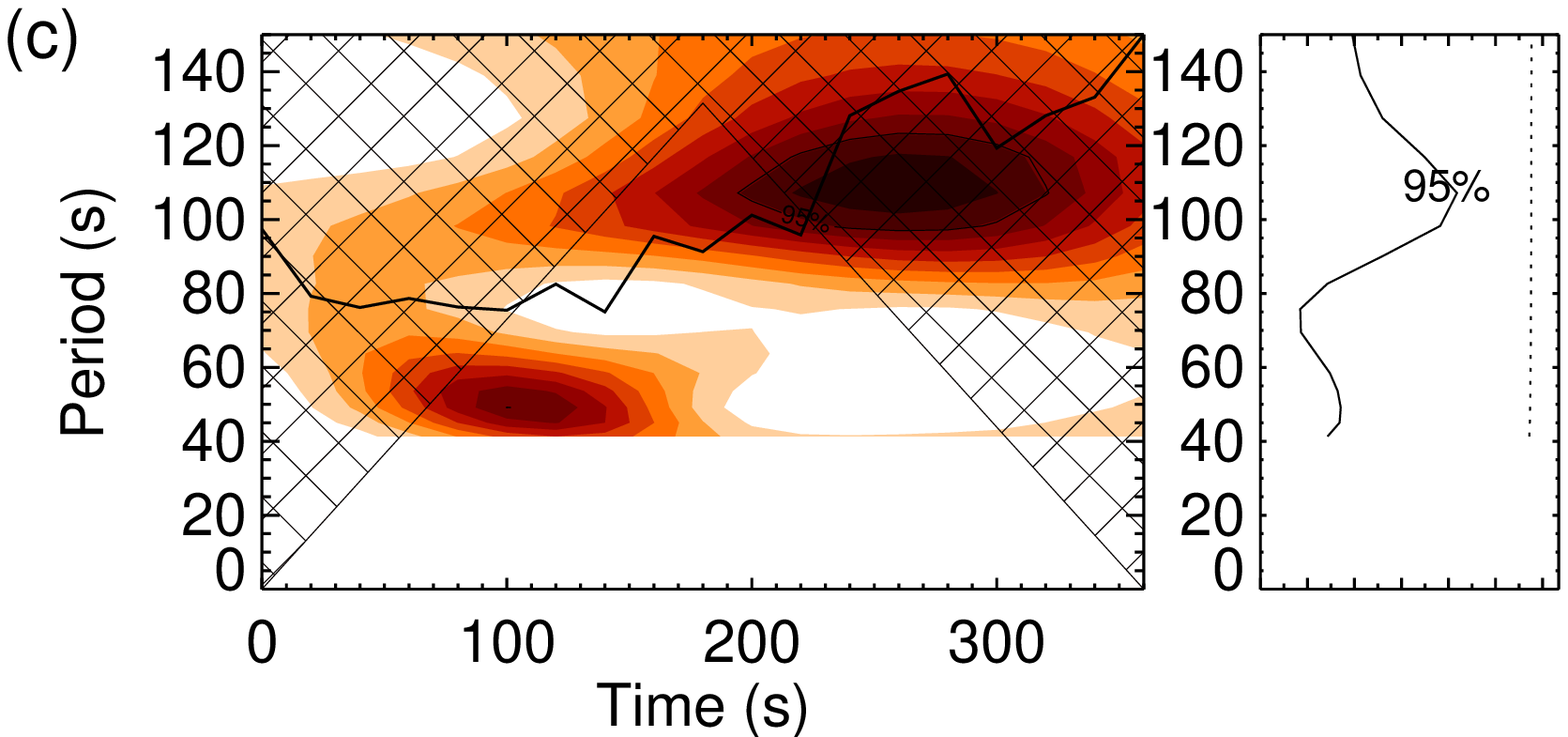}
		}
		\caption{
			Wavelet power spectra of the following time profiles:
			(a) temperature, 
			(b) emission measure, 
			(c) errors of the spectral fitting.
			The time axis corresponds to the interval 17:53:04--17:58:11~UT. The cadence time is $\Delta t = 20$~s. See Section~\ref{s:Periods} for a detailed description.}
		\label{f:wavelet2}
	\end{figure*}

	The periodic properties of the time series of the temperature and emission measure are presented in Figure~\ref{f:wavelet2}.
	Each panel in Figure~\ref{f:wavelet1} and Figure~\ref{f:wavelet2} contains the wavelet power spectrum (colored plot) and global wavelet spectrum (to the right of colored plot) of the high-frequency component of different data. We have fixed the width of the smoothing interval to $\tau = 100$~s. In each panel, the normalized time series is over-plotted in the wavelet power spectrum. The green contour indicates the 95\% significance level. The red noise background spectrum was chosen for the calculation of the significance level. The plot to the right of the colored one is the global wavelet spectrum obtained by integration of the wavelet power spectrum over time. The dashed line here shows the 95\% significance level assuming the more strict red noise. 
	
	The error of the period is commonly defined by both the Fourier frequency sampling and the width of the spectral peak in the periodogram or wavelet spectrum. The frequency sampling is equidistant in contrast to the period sampling, which is $P = 1/f$. For the period, the error bars are not equal at both sides of the spectral peak. In the paper, we thus use both error values for the periods. So, we obtain a period $P \approx 72^{+6.5}_{-5.9}$~s, where the errors are determined by the Fourier frequency grid. The uncertainty in the period caused by the width of the spectral peak in the wavelet spectrum is $\Delta P_\mathrm{HWHM} \approx 10$~s.
	
	A similar period $P \approx 70$--$75$~s is detected in the microwave emission at frequency 15.7~GHz. The fact that a similar periodicity is found in different time series with different instruments provides evidence in favor of this periodicity not being artificial, and that it relates to a solar phenomenon. The spectral peak in the global wavelet spectrum is wider than that obtained for the HXR emission resulting in an error of $\Delta P_\mathrm{HWHM} \approx 13$~s. 
	
	Moreover, time profiles of the temperature and emission measure (see Section~\ref{s:ObservationsXRays}) demonstrate the same periodicity, $P \approx 74$--80~s (Figure~\ref{f:wavelet2}). Here we have used a cadence time of $\Delta t = 20$~s, which we chose in order to increase the signal-to-noise ratio. This time resolution is high enough to resolve the 80-s periodicity. The uncertainties of the period are $\Delta P_\mathrm{FFT} \approx 12$~s and $\Delta P_\mathrm{HWHM} \approx 15$~s.
	
	Note that the shortest period in the vertical axis in Figure~\ref{f:wavelet1} and Figure~\ref{f:wavelet2} is defined as double time resolution, $2 \Delta t$. In Figure~\ref{f:wavelet1} we use data with $\Delta t = 4$~s. Therefore,  the shortest period, that can be defined by wavelet transform, equals to 8~s. In Figure~\ref{f:wavelet2}, $\Delta t = 20$~s causes the shortest period to be equal to 40~s, leaving  blank the part of the wavelet spectrum below that value.
	
	\subsection{Phases}\label{s:Phases}
\begin{figure*}   
		\centerline{
			\includegraphics[width=0.5\textwidth,clip=]{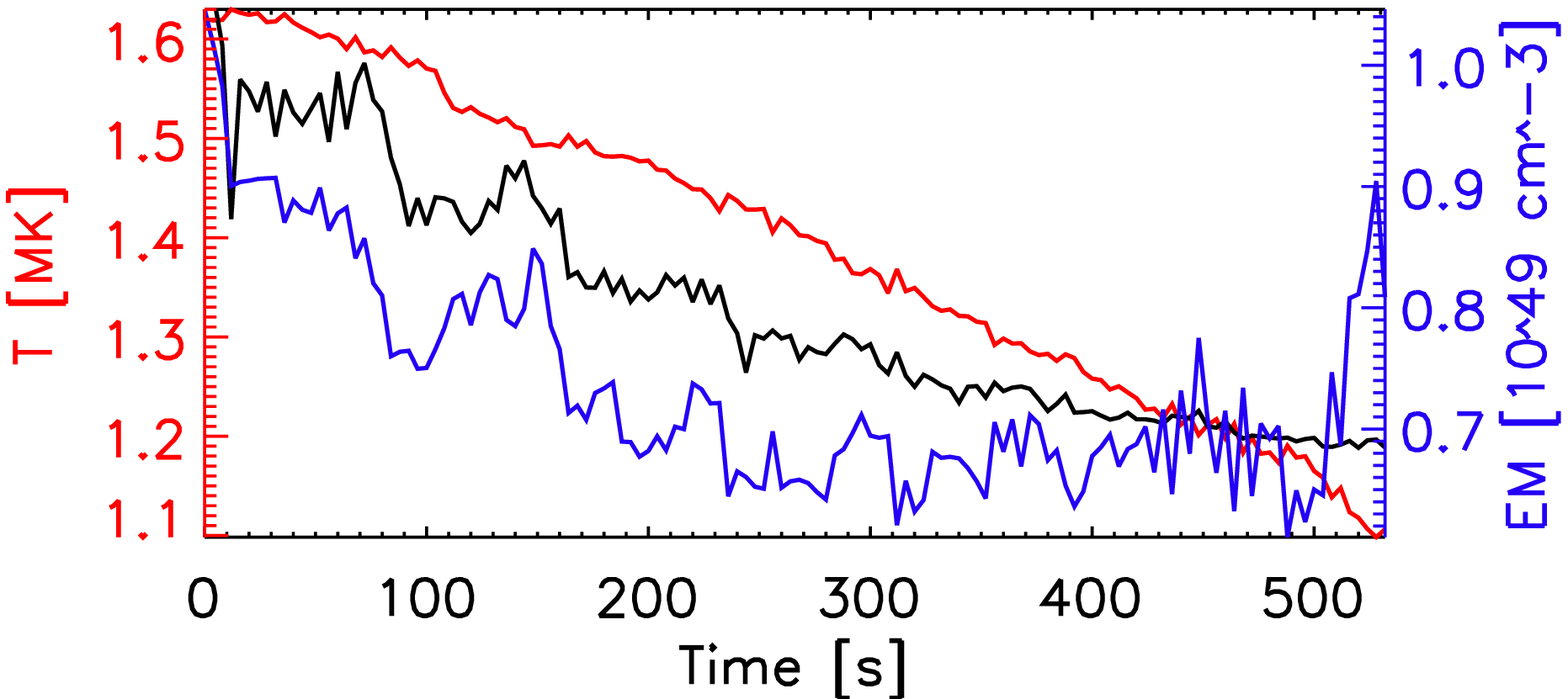}
		}
		\centerline{
			\includegraphics[width=0.420\textwidth,clip=]{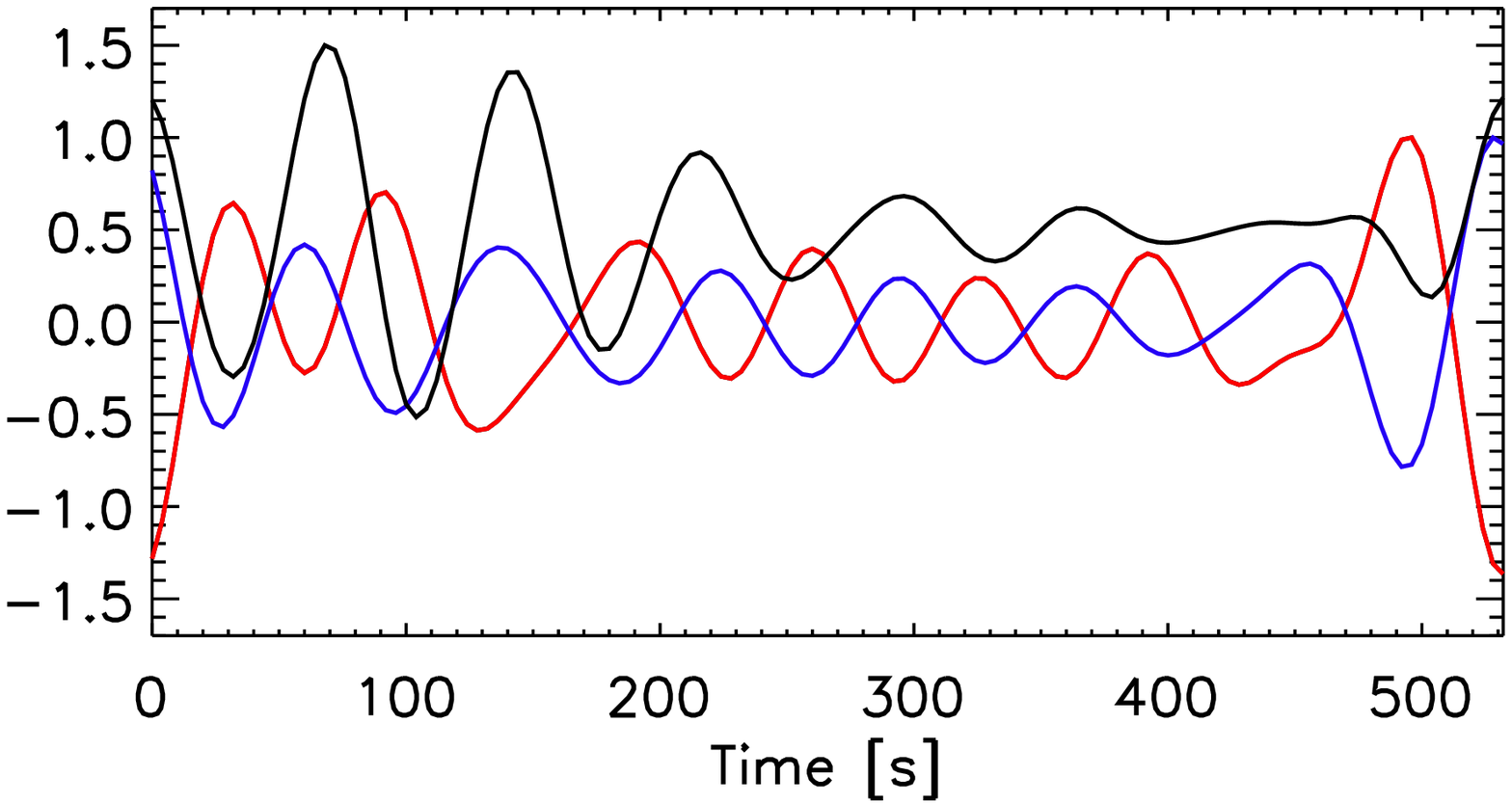}
			\includegraphics[width=0.420\textwidth,clip=]{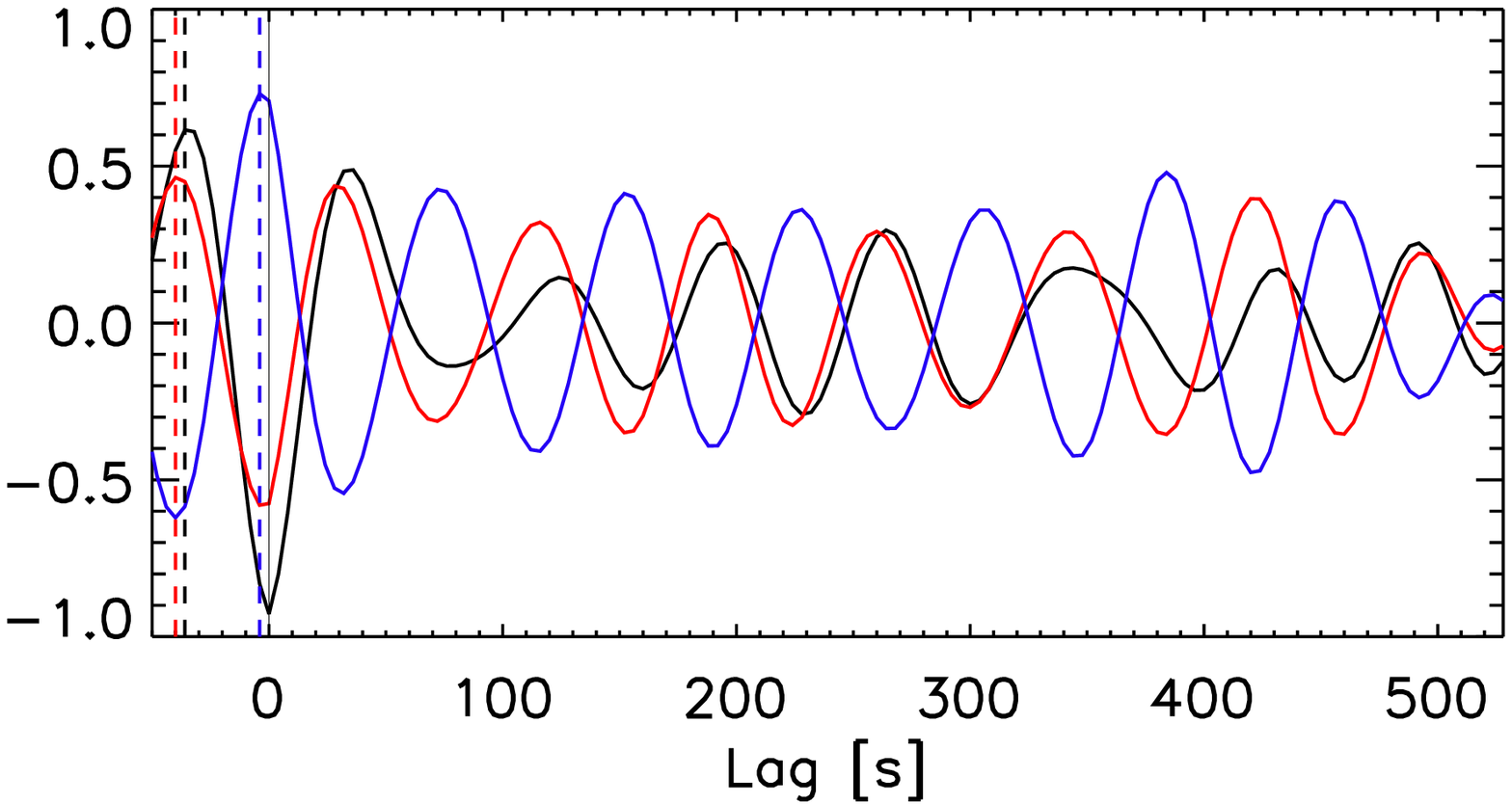}
		}
		\caption{
			Upper panel:  the total time series of temperature $T_e$ (red), emission measure $EM$ (blue) and flux at 12--25~keV (black). The cadence time is $\Delta t = 4$~s.
			Lower left panel: the high-frequency component of the time profiles of  $T_e$ (red), $EM$ (blue) and flux at 12--25~keV (black). Lower right panel: cross-correlation functions between $EM$ and $T_e$ (black), between $T_e$ and flux at 12--25~keV (red) and between $EM$ and flux at 12--25~keV (blue). The colored vertical dashed lines indicate the maxima of the corresponding cross-correlation functions.}
		\label{f:TeEM_ccorr}
	\end{figure*}

	The analysis of phase relationships between the time profiles of the temperature, emission measure and flux at 12--25~keV was performed. The total (with trend) time series of temperature, emission measure and flux at 12--25~keV derived from X-ray RHESSI data using the imaging spectroscopy method (see upper plot at Figure~\ref{f:TeEM_ccorr}) are plotted in the upper panel. Here, the time profiles of $T_e$ and $EM$ are shown in their absolute values while the time profile of the flux is normalized over the maximum of $EM$. The cadence time is $\Delta t = 4$~s. The high-frequency components of the time profiles are shown in the bottom-left panel. The cross-correlation functions between $EM$ and $T_e$, between $T_e$ and flux and between $EM$ and flux are shown in the bottom-right panel. Taking into account that the uncertainty is about the cadence time $\Delta t = 4$~s, we could assert the in-phase behavior of the oscillations of the flux and emission measure. Two other cross-correlation functions show anti-phase behavior with the phase shift $\Delta \varphi \approx 36$--$40$~s the oscillation of $T_e$ relatively to the oscillations of $EM$. A similar relation is also obtained from the Fourier transform.
	
\subsection{Spatial features in X-rays and EUV related with QPPs}
\begin{figure*}   
	\centerline{
		\includegraphics[width=0.43\textwidth,clip=]{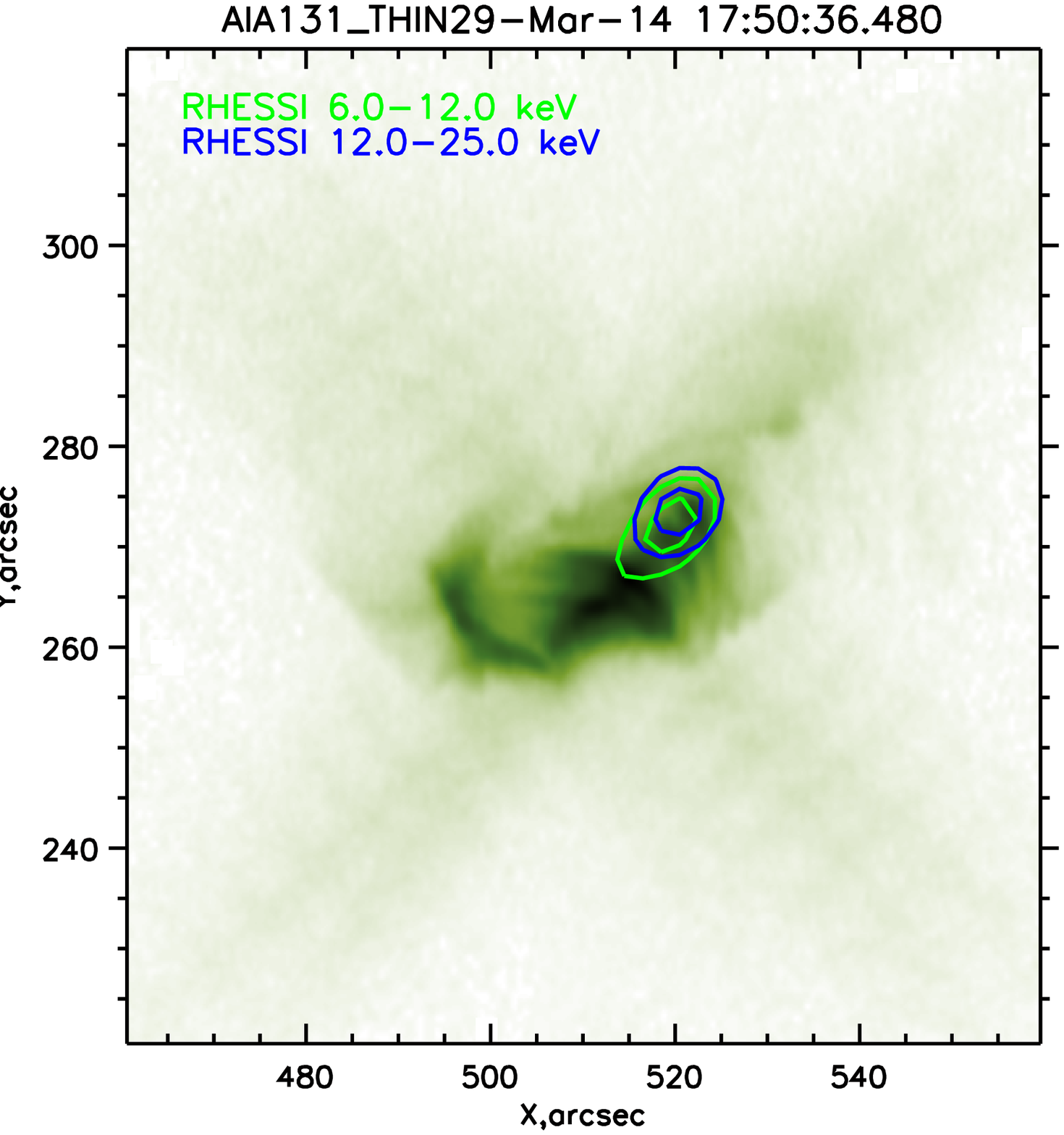}
		\includegraphics[width=0.48\textwidth,clip=]{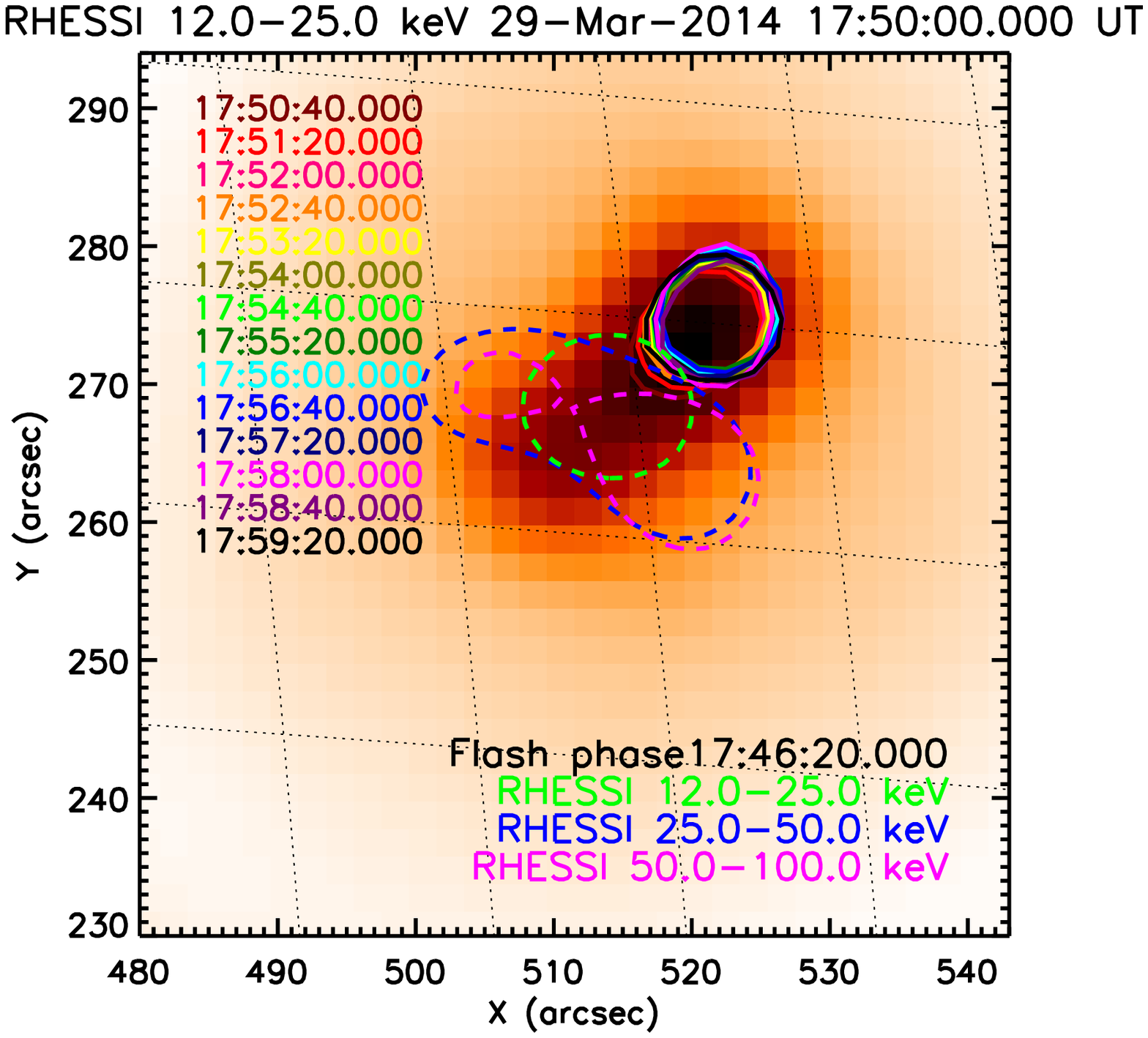}
	}
	\caption{
		Left panel: positions of the hard X-ray sources (green and blue contours) relative to the loop arcade in SDO/AIA 131{\AA} emission (background image).
		Right panel: the dynamics of the HXR source at 12--25~keV through the decay phase is shown by the solid colored contours. The dashed contours correspond to the emission at flare maximum at 12--25~keV (green contours), 25--50~keV (blue contours), and 50--100~keV (pink contours). The background image shows 12--25~keV emission at the beginning of the decay phase (15:50:40~UT).}
	\label{f:SDO_hsi_images}
\end{figure*}

	X-ray emission at the flash (or impulsive) phase clearly shows the loop top source at 12--25~keV and two footpoints at 50--100~keV (dashed contours in Figure~\ref{f:SDO_hsi_images}, right panel). 
	
	During the decay  phase of the flare related with the QPPs, we were able to localize the source of X-ray emission within 12--25~keV energy band only. The source located on the top of the loop arcade clearly seen in EUV emission (see Figure~\ref{f:SDO_hsi_images}, left panel).  
	
	Analysis of the dynamics of the source at 12--25~keV reveals that the source remains stable during the whole decay phase (solid contours in Figure~\ref{f:SDO_hsi_images}, right panel). Note that the X-ray source keeps a stable position during the decay phase. Moreover, its size does not change through the decay phase.  
	
	Using EUV images, we estimate size of the arcade: the length of the loops in the arcade is $L \approx 17$--$22$~Mm and the length of the arcade is about $10$--$12$~Mm.

	\vspace{0.6cm}
	Summing up the results of the study done in Section~\ref{s:QPP}, we emphasize the following features of the QPPs. 
	\begin{enumerate}
		\item The period of QPPs is $P \approx 80$~s and it is similar in both X-rays and microwaves. The fact that the properties of the QPPs are similar in different wavelength ranges proves that the QPPs are real, and they are not an artefacts of the data handling. 
		\item On the other hand, similar QPPs are observed by both ground based radio telescope and X-ray satellite. This excludes the possibility of the QPPs as an effect of the atmosphere of the Earth. 
		\item Moreover, QPPs with the same period are found in the time profiles of the temperature and emission measure with the anti-phase behavior of those time profiles. 
		\item The source of the X-ray emission in which the QPPs were found  is located at the top of the small almost uniform loop arcade.
	\end{enumerate}
	What model explains the features of the QPPs on the decay phase? In the next section we discuss the scenarios which would explain the periods and unusual phases of the QPPs during the decay phase.
	
	\section{Interpretation}\label{s:Mechanisms}
	
	\subsection{Interpretation of the period}\label{s:Reasons_Period}
	We found QPPs with periods $P \approx 74$--80~s in different data: in X-rays at energies of 3--25~keV, in microwaves at 15.7~GHz, and in both the temperature and emission measure. The quality of the oscillations is rather low, each time profile contains from 5 to 6 periods before damping to the noise level.
	
	The variation of the emission measure found in our study implies a variation of either the flaring volume or the plasma density.  The kink mode is weakly compressible resulting in density (temperature) perturbations of about 0.3--0.4\% of the equilibrium value \citep{2008ApJ...676L..73V,2012ApJ...753..111G, 2008A&A...485..849V}.  Variations of a volume are also negligibly small. Moreover, for kink waves, it is expected that the $EM$ and $T_e$ are in phase \citep{2016ApJS..223...23Y}. Thus, multiple waves should be excited to have the observed, almost anti-phase, behavior, and this is highly unlikely. 
	
	So, we consider slow magneto-acoustic (SMA) and fast sausage modes. Both modes are compressible and could modulate plasma density (SMA mode, sausage mode) or volume (sausage mode) producing a signature in the $EM$ \citep{2013A&A...555A..74A,2014ApJ...785...86R,2015ApJ...807...98Y}. 
	However, it is difficult to associate the observed QPPs with the sausage mode because its period is usually of the order of a few seconds, i.e. much shorter than the observed periods. We use the following values obtained in Section~\ref{s:Periods} and Section~\ref{s:Phases} to estimate the periods of an MHD mode: loop length $L \approx 17$--$22$~Mm, temperature $T_e \approx 1.1$--$1.6$~MK. Using $EM \approx 0.6$--$0.9 \times 10^{49}$~cm$^{-3}$ and the linear source size of $10$'' we estimate a plasma density $N \approx 1.2$--$1.5\times 10^{11}$~cm$^{-3}$. The assumed magnetic field is $B \approx 100$--$150$~G which is reasonable for a small loop arcade. For the given configuration, a phase speed of 170--200~km/s is estimated using the dispersion equation \citep{2007AdSpR..39.1804N}. 
	The corresponding period of the global standing SMA mode is too long to explain the observed period. On the contrary, period of the second harmonic of the standing SMA mode is estimated as $P \approx 73$--$95$~s. This value matches much better the observed period. Moreover, the enhanced compression caused by the second harmonic occurs at the footpoints  and, especially, at the loop top. This should result in that the maximal flux variations are expected to be at the loop top. This facilitates the detection of the QPPs in the X-ray source observed at the top of arcade (see Figure~\ref{f:SDO_hsi_images}). Unfortunately, we can not check the variations in the footpoints because they are not pronounced in X-rays.
	Thus, we conclude that second harmonic of the standing slow magneto-acoustic mode is the most likely explanation for the observed periods.

	\subsection{Interpretation of the phase shift}\label{s:Reasons_Phase}
	The damping oscillations in the emission measure and temperature have an unusual anti-phase behavior with the phase shift is about $\Delta \varphi \approx 40$~s (Figure~\ref{f:TeEM_ccorr}). In this Section we discuss different scenarios for these features.

	\subsubsection{Non-ideal MHD}\label{s:NonidealMHD}
	A phase is introduced between the density and temperature, when thermal conduction (or other non-ideal effects) is introduced when modelling slow magneto-acoustic waves. On the one hand, the density is ahead of the temperature for the standing slow waves \citep{2009A&A...494..339O, 2016ApJ...820...13M}. On the other hand, the density is behind the temperature for thermal conduction fronts \citep{2015ApJ...813...33F}. 
	However, having a half-period phase shift $\Delta \varphi \approx P/2 = \pi$, it is difficult to decide which oscillations are started first. Note that the work by \citet{2018arXiv181008449K} also finds a phase shift of $\pi$ between temperature and emission measure for slow waves in fan loops.
	
	Previously, the effect of the phase delay has been used to measure the thermal conduction with propogating slow waves in coronal loops \citep{2011ApJ...727L..32V, 2015ApJ...811L..13W}. Therefore, the phase shift between the density and temperature is compatible with such a model of a standing slow wave including thermal conduction. Using the formulae in the previous papers, we can estimate the value for the thermal conduction coefficient. 
	Using estimates for the temperature $T_e \approx 1.1$--$1.6$~MK (see Section~\ref{s:Phases}) we expect the Spitzer thermal conductivity parallel to the magnetic field equals to $k_\parallel = 7.8 \ 10^{-7} T^{5/2} \approx 2.2 \ 10^5$--$4.6 \ 10^5$~erg cm$^{-1}$ s$^{-1}$ K$^{-1}$.
	Unfortunately, in the case of the anti-phase oscillations, we can not check if the observed phase shift $\Delta \varphi$ corresponds well to the estimated thermal conductivity because Equation (4) in \citep{2015ApJ...811L..13W} is true for $\Delta \varphi$ from $0$ to $\pi/2$ only. 
	
	\subsubsection{Non-linearity}\label{s:Nonlinear}
	Another possibility for causing the phase delay could be non-linear behavior of the slow waves, that could lead to non-intuitive behavior for density and temperature. It is expected that the waves are non-linear, because the amplitude of the waves is very large in flares. Recently, the non-linear behavior of slow waves was modeled by e.g. \citet{2017ApJ...849...62N}, but they did not focus on the phase delay between density and temperature in the non-linear regime. However, non-linear behavior of the wave would also result in non-sinusoidal variation of the intensity, density and temperature. This would result in the detection of higher harmonics in the signal \citep{2018RvMPP...2....2D}, but this is not seen in our observations. 
	
	\subsubsection{Multiple loops}\label{s:Multiple}
	It is also possible that the phase shift is caused by the intensity variations in different subsequent loops in the flaring arcade. This would be compatible with the observed Morton wave \citep{2016SoPh..291.3217F}, which could have sequentially perturbed the loops in the arcade. The phase shift could then be determined by the periodic enhancing of the intensity of different loops, resulting in this peculiar phase shift. However, the question then is how the period remains so stable. If the sound speed is slightly different in two loops, then the oscillations would go out of phase very rapidly destroying the coherence between two loops. Still, it may work with an arcade, because the loops are probably more or less equally long and their temperature is not too different either. The coherent emission between subsequent loops would also explain the apparent slow cooling of the loop, which is happening on a much longer time scale than the thermal conduction time scale. One loop would rapidly cool down and disappear from the passband, but while it cools down a new one (almost co-spatial) would cool into the passband as well, significantly lengthening the time the ``loop'' is visible in the passband. This scenario implies the progression of the X-ray source along the axis of the arcade, towards the North-West. However, the drift is not pronounced in images (Figure~\ref{f:SDO_hsi_images}, right panel). Probably, the spatial resolution of RHESSI is too low to detect the drift across the small arcade.
	
	\subsubsection{Additional heating processes}\label{s:Heating}
	We could also consider the Moreton wave as an indirect reason of the QPPs.  Analyzing this event, \citet{2016SoPh..291.3217F} found both the Moreton wave which was triggered by the flare and its counterpart in the corona in EUV emission. 
	The speed of the wave in the chromosphere $v_{M_{chr}} \approx 461$--715~km/s in these sectors. The corresponding speed in the corona is $v_{M_{cor}} \approx 956$--1240~km/s. Using the average value the wave speed $\overline{v}_{M_{chr}} \approx 588$~km/s and the length of the arcade 11~Mm (see Section~\ref{s:Observations}) one finds that the wave passes through the arcade in 19~s. This time is four times shorter than the period of the QPPs. Moreover, having such speed the Moreton wave passes through the arcade much earlier than the oscillations of $T_e$ and $EM$ become pronounced. So, the Moreton wave can not be the direct cause of the QPPs. But, propagating along the arcade axis the Moreton wave could 
	perturb plasma
	in the loop footpoints simultaneously during the impulsive flare phase. This may trigger the second harmonic of the standing slow magneto-acoustic mode in the loops \citep{2004A&A...414L..25N, 2004A&A...422..351T, 2015ApJ...804....4K}. 
    \citet{2011ApJ...730L..27N} suggested a model in which a slow magneto-acoustic wave may trigger reconnection in the current sheet above the arcade. The reconnection leads to multiple acts of energy release and therefore increasing plasma temperature with subsequent increase of emission measure. 
	
	\section{Conclusions}\label{s:Conclusions}
	
	Based on the analysis presented in this paper, quasi-periodic pulsations with a period $P \approx 80$~s are found in the microwave and X-ray emissions during the decay phase of the solar flare SOL2014-03-29T17:48.  The similarity of the periods in both X-rays registered by spacecraft and microwaves registered by on-ground radio telescope proves that the QPPs are neither an artifact of data handling nor an effect of the ionosphere of the Earth. Moreover, temperature and density oscillate with the same period. We found an unusual anti-phase behavior of the temperature oscillations and density oscillations. We postulate that the observed pulsations best correspond to the second harmonic of slow magneto-acoustic mode in each loop of the small arcade triggered by a Moreton wave, which increased the plasma density in the two footpoints simultaneously. The slow magneto-acoustic waves trigger reconnection in the current sheet above the arcade. The slow propagation of the reconnection along the arcade leads to additional quasi-periodic energy release and plasma heating.
	
	\section*{Acknowledgements}
	
	This research is partly supported by the following grants of the Russian Foundation for Basic Research (RFBR): RFBR No. 17-52-80064~--- analysis of high-energy flare, RFBR No. 17-52-10001~--- chapter~\ref{s:QPP} related to the technique of analysis of the QPPs; No.18-02-00856~--- chapter~\ref{s:Mechanisms}. LKK  thanks the budgetary funding of Basic Research program II.16. TVD was supported by GOA-2015-014 (KU~Leuven). EGK thanks a mobility grant from Belspo and the RAS Presidium program No. 28. PC acknowledges BK21 plus program of the National Research Foundation (NRF) funded by the Ministry of Education of Korea. This work was based on discussions at the ISSI and ISSI-Beijing. This project has received funding from the European Research Council (ERC) under the European Union's Horizon 2020 research and innovation programme (grant agreement No 724326). AKS acknowledges the RESPOND-ISRO (DOS/PAOGsIA2015-16/130/602) grant, and Indo-US (IUSSTF) Joint R\&D Networked Center (IUSSTF-JC-011-2016) for the support of his research.  
	
	
	\bibliographystyle{mnras}
	\bibliography{references} 
	
	
	\bsp	
	\label{lastpage}
\end{document}